\documentclass[sn-mathphys,Numbered]{sn-jnl}

\usepackage{graphicx}%
\usepackage{multirow}%
\usepackage{amsmath,amssymb,amsfonts}%
\usepackage{amsthm}%
\usepackage{mathrsfs}%
\usepackage[title]{appendix}%
\usepackage{xcolor}%
\usepackage{textcomp}%
\usepackage{manyfoot}%
\usepackage{booktabs}%
\usepackage{listings}%

\theoremstyle{thmstyleone}%
\theoremstyle{thmstyletwo}%

\theoremstyle{thmstylethree}%

\newcommand{\ket}[1]{\left|#1\right>}
\newcommand{\bra}[1]{\left<#1\right|}

\newcommand{\nn}{\nonumber\\}

\newcommand{\f}[1]{\mbox{\boldmath$#1$}}

\newcommand{\bea}{\begin{eqnarray}}
\newcommand{\ea}{\end{eqnarray}}
\newcommand{\eea}{\end{eqnarray}}
\newcommand{\ord}{\,{\cal O}}

\newcommand{\ii}{{\rm i}}

\begin{document}

\title[Attraction versus repulsion between doublons or holons in Mott-Hubbard systems]{Attraction versus repulsion between doublons or holons in Mott-Hubbard systems}


\author*[1,2]{\fnm{Friedemann} \sur{Queisser}}\email{f.queisser@hzdr.de}

\author[1]{\fnm{Gernot} \sur{Schaller}}

\author[1,2]{\fnm{Ralf} \sur{Sch\"utzhold}}

\affil[1]{Helmholtz-Zentrum Dresden-Rossendorf, 
Bautzner Landstra{\ss}e 400, 01328 Dresden, Germany}

\affil[2]{Institut f\"ur Theoretische Physik, 
Technische Universit\"at Dresden, 01062 Dresden, Germany}


\abstract{For the Mott insulator state of the Fermi-Hubbard model 
in the strong-coupling
limit, we study the interaction between quasi-particles in the form of doublons 
and holons. 
Comparing different methods -- the hierarchy of correlations, 
strong-coupling perturbation theory, and exact analytic solutions for the 
Hubbard tetramer -- we find an effective interaction between doublons and/or 
holons to linear order in the hopping strength 
which can display 
attractive as well as repulsive contributions, 
depending on the involved momenta.  
%
%
Finally, we speculate about the implications of our findings for 
high-temperature superconductivity. }




\maketitle

\section{Introduction}\label{sec1}

Understanding strongly interacting quantum many-body systems is one of the 
major challenges of contemporary physics.
In order to achieve progress in that direction, it is often useful to apply 
the same principles as for weakly interacting systems. 
Following this strategy, the first step is to find or characterize the ground 
or thermal equilibrium state \cite{Imada89,Qin20,LeBlanc15}.
As the second step, one should identify the relevant quasi-particle excitations 
describing linearized perturbations around this equilibrium state and determine 
their properties, such as dispersion relations \cite{Herr97,Kung15,Ede90,Belk95,Voj98,Ble22}. 
Going beyond this linearized level, one can then study the interactions of 
these quasi-particle excitations among each other and with other degrees of 
freedom \cite{Kuz94,Bul93,Cher94,Bel95,Bel97,Shr88,Poi94,BohARXIV,Gru23,Boh22,Bar89}. 

Since exact solutions are typically limited to special cases or small systems \cite{LiebWu,Essler,Krut16}, 
approximations are necessary in most cases. 
Ideally, these approximation schemes should be based, at least in principle, 
on a systematic expansion into powers of some small control parameter.
In contrast to weakly interacting systems, the large coupling strength 
prohibits its use as perturbation parameter, 
but one could use its inverse \cite{Pair98,Rohr18,Sen02}
(strong-coupling perturbation theory) or the inverse of some other large number, 
such as spin $S\gg1$ \cite{Hol40,Iga05,Ogu60} or coordination number $Z\gg1$ 
\cite{Br59,Nav10,Krut14,Queiss19,Queiss14,Nav16}, which typically leads to 
some sort of mean-field theory. 

In the following, we consider the Fermi-Hubbard model as the {\em drosophila} 
of strongly interacting quantum many-body systems \cite{Hub63,Aro22,Qin22} ($\hbar=1$)
\bea
\label{Fermi-Hubbard}
\hat H=-\frac1Z\sum_{\mu\nu s} T_{\mu\nu} \hat c_{\mu s}^\dagger \hat c_{\nu s} 
+U\sum_\mu \hat n_\mu^\uparrow\hat n_\mu^\downarrow
\,,
\ea
where $\hat c_{\mu s}^\dagger$ and $\hat c_{\nu s}$ denote the fermionic 
creation and annihilation operators at the lattice sites $\mu$ and $\nu$ 
with spin $s\in\{\uparrow,\downarrow\}$ while $\hat n_\nu^s$ are the 
associated number operators. 
The lattice structure is encoded in the hopping matrix $T_{\mu\nu}$
which equals the tunneling strength $T$ for nearest neighbors $\mu$ and $\nu$ 
and is zero otherwise. 
The coordination number $Z$ counts the number of nearest neighbors $\mu$
for a given lattice site $\nu$ and is assumed to be large $Z\gg1$.
Finally, $U$ denotes the on-site repulsion and we focus on 
the strong-coupling limit $U\gg T$ in the following. 

Let us briefly recapitulate the relevant symmetries of the Fermi-Hubbard
Hamiltonian~\eqref{Fermi-Hubbard}. 
In addition to the total particle numbers $\hat N^s=\sum_\mu \hat n_\mu^s$, 
the total spin 
\bea
\label{spin}
\hat{\mathbf{S}}
=
\sum_\mu\hat{\mathbf{S}}_\mu
=
\frac12
\sum_{\mu ss'}\f{\sigma}_{ss'}\hat c^\dagger_{\mu s}\hat c_{\mu s'}
\ea
is also conserved, where $\f{\sigma}_{ss'}$ are the elements of the 
Pauli spin matrices reflecting the global $SU(2)$-invariance \cite{Lieb89}. 
Here, bold-face symbols such as 
$\hat{\mathbf{S}}=(\hat S^x,\hat S^y,\hat S^z)$ represent vectors.

Another interesting symmetry is the particle-hole duality:
If we exchange all creation and annihilation operators 
$\hat c_{\mu s}^\dagger \leftrightarrow \hat c_{\mu s}$
which implies 
$\hat n_\mu^s \leftrightarrow 1-\hat n_\mu^s$, 
we find that the Hamiltonian~\eqref{Fermi-Hubbard} 
is mapped to the same form with a negative hopping strength 
$T \leftrightarrow -T$ up to an irrelevant shift containing 
the total particle number $\hat N=\hat N^\uparrow+\hat N^\downarrow$. 
In order to avoid this shift, one could consider the 
grand-canonical Hamiltonian $\hat H_{\rm gc}=\hat H-\mu\hat N$
with the chemical potential $\mu=U/2$ which is then mapped onto itself
with $T \leftrightarrow -T$.

For bi-partite lattices, where one can introduce a parity $(-1)^\mu$ 
which is alternating for neighboring lattice sites, the pseudo-spin 
$\hat{\f{\eta}}=(\hat\eta^x,\hat\eta^y,\hat\eta^z)$ 
leads to another conserved quantity 
(see Appendix~\ref{APP:pseudospin}). 
In addition, for this case the staggered gauge transformation 
$\hat c_{\mu s}\to(-1)^\mu\hat c_{\mu s}$ does also map the 
Hamiltonian~\eqref{Fermi-Hubbard} into the same form with a 
negative hopping strength $T \leftrightarrow -T$ \cite{Yang89,Zhang90,Essler}. 

A prominent example for the crucial differences between weakly and strongly 
interacting systems is the Mott insulator \cite{Hub63,Avi20}.
For weak interactions $U\ll T$, the state at half filling for both spin 
species would be metallic, only the Fermi surface would be deformed a bit 
by the coupling $U$.
The Mott insulator \cite{Ima98} is realized in the other limit $U \gg T$ 
however, where
the ground state is insulating and 
basically one particle occupies each lattice site -- up to small virtual 
hopping corrections with probabilities of order $T^2/U^2$. 
In order to facilitate transport, one has to excite a doublon-holon pair 
which requires a minimum energy given by the Mott gap 
$\Delta E_{\rm Mott}\approx U$. 
Note that, in contrast to these real and long-lived doublon-holon pairs 
(whose creation requires a minimum energy given by the Mott gap 
$\Delta E_{\rm Mott}\approx U$), the hopping corrections mentioned above 
are sometimes pictured as virtual and short-lived doublon-holon pairs 
(which do not require such an excitation energy and are present in 
the ground state).

In the following, we shall study the properties of these quasi-particle 
excitations on top of the Mott insulating state \cite{Pre95}. 
As explained above, this includes the single-particle characteristics 
such as their dispersion relation -- but also two-particle properties
describing their interaction among each other.

\section{Hierarchy of Correlations}  

In order to pursue the strategy described in the Introduction, let us first 
employ the method of the hierarchy of correlations, see also 
\cite{Nav10,Queiss19b,Queiss14}. 
To this end, we consider the reduced density matrices of one $\hat\rho_\mu$, 
two $\hat\rho_{\mu\nu}$, and three $\hat\rho_{\mu\nu\lambda}$ lattice sites, 
etc., and split up the correlated parts via 
$\hat\rho_{\mu\nu}^{\rm corr}=\hat\rho_{\mu\nu}-\hat\rho_{\mu}\hat\rho_{\nu}$, 
and so on. 

Now, based on the assumption $Z\gg1$, we may employ an expansion into powers 
of $1/Z$ where we find that higher-order correlators are successively
suppressed as $\hat\rho_{\mu\nu}^{\rm corr}=\ord(1/Z)$,
$\hat\rho_{\mu\nu\lambda}^{\rm corr}=\ord(1/Z^2)$, and so on.  
This hierarchy facilitates an iterative approximation scheme, 
where we may start from the exact evolution equations
\bea
\label{evolution}
i\partial_t \hat\rho_\mu 
&=& 
F_1(\hat\rho_\mu,\hat\rho_{\mu\nu}^{\rm corr})
\,,\nn
i\partial_t \hat\rho_{\mu\nu}^{\rm corr} 
&=& 
F_2(\hat\rho_\mu,\hat\rho_{\mu\nu}^{\rm corr},\hat\rho_{\mu\nu\lambda}^{\rm corr})
\,,\nn
i\partial_t \hat\rho_{\mu\nu\lambda}^{\rm corr} 
&=& 
F_3(\hat\rho_\mu,\hat\rho_{\mu\nu}^{\rm corr},\hat\rho_{\mu\nu\lambda}^{\rm corr},
\hat\rho_{\mu\nu\lambda\kappa}^{\rm corr})
\,,
\ea
and so on for even higher orders, where the functions $F_n$ are determined by 
the Hamiltonian~\eqref{Fermi-Hubbard}.  

To lowest order $\ord(Z^0)$, we may approximate the first equation by 
$i\partial_t \hat\rho_\mu=F_1(\hat\rho_\mu,0)+\ord(1/Z)$. 
The solution to this equation obeying the required boundary conditions 
then yields the mean-field ansatz $\hat\rho_\mu^0$ as the starting point 
for calculating the higher orders in $1/Z$.


In order to describe the Mott insulator state at half filling in the 
strong-coupling limit $U\gg T$, we use the simple mean-field ansatz 
at zero temperature)
\bea
\label{mean-field}
\hat\rho_\mu^0
=
\frac{\ket{\uparrow}_\mu\!\bra{\uparrow}+\ket{\downarrow}_\mu\!\bra{\downarrow}}{2}
\,.
\ea
In principle, the aforementioned virtual hopping corrections with small 
probabilities $\sim T^2/U^2$ for an empty $\ket{0}_\mu$ or full lattice site 
$\ket{\uparrow\downarrow}_\mu$ could be included as well, 
but we neglect them here. 

Note that the above mean-field ansatz~\eqref{mean-field} 
is invariant under the particle-hole duality
transformation mentioned after Eq.~\eqref{spin} and
does 
not include any spin ordering to lowest order. 
A staggered mean-field ansatz which does display spin ordering could 
be introduced for bi-partite lattices via 
\bea
\label{Ising}
\hat\rho_\mu^{\rm Ising}
=
\left\{
\begin{array}{ccc}
\ket{\uparrow}_\mu\!\bra{\uparrow} & {\rm for} & \mu\in\mathcal A
\\
\ket{\downarrow}_\mu\!\bra{\downarrow} & {\rm for} & \mu\in\mathcal B
\end{array}
\right. 
\,,
\ea
where $\mathcal A$ and $\mathcal B$ denote the two sub-lattices. 
This state describes an Ising type anti-ferromagnet where 
$\langle\hat{S}^z_\mu\hat{S}^z_\nu\rangle$ is minimized and the 
${\mathbb Z}_2$ symmetry 
$\hat c_{\mu\uparrow} \leftrightarrow \hat c_{\mu\downarrow}$
is spontaneously broken. 

Note, however, that the ground state of the Fermi-Hubbard 
model~\eqref{Fermi-Hubbard} does not display this Ising type 
but rather Heisenberg type anti-ferromagnetic 
order where $\langle\hat{\mathbf{S}}_\mu\cdot\hat{\mathbf{S}}_\nu\rangle$ 
is minimized instead of $\langle\hat{S}^z_\mu\hat{S}^z_\nu\rangle$ 
as in the Ising case. 
This ground state is invariant under the $SU(2)$-invariance generated 
by the total spin~\eqref{spin} instead of the broken ${\mathbb Z}_2$ 
symmetry of the Ising case. 
Formally, 
the reduced density matrix of a single lattice site $\mu$ 
is given by the ansatz~\eqref{mean-field}. 
The correlations between neighboring lattice sites $\mu$ and $\nu$ 
can be taken into account via $\hat\rho_{\mu\nu}^{\rm corr}$.
As a more intuitive picture, the Heisenberg type anti-ferromagnet 
can be visualized as lying somewhere in between the fully ordered 
Ising-type state~\eqref{Ising} and the state~\eqref{mean-field} 
without any spin order, see also Eq.~\eqref{Mott-zero} below.

The correlations $\hat\rho_{\mu\nu}^{\rm corr}$ can be further suppressed 
for finite temperatures. 
For example, if the temperature is much larger than the effective 
anti-ferromagnetic interaction $\ord(T^2/U)$ but still way below the Mott gap
$\Delta E=\ord(U)$, the ansatz~\eqref{mean-field} would basically 
reproduce the exact thermal density matrix.
As another possibility, the coupling to an environment can effectively steer 
the system towards the state~\eqref{mean-field}, see, e.g., \cite{Queiss19,Sch22}.

%
%
%

\subsection{Doublons and holons}  

To next order in $1/Z$, we may derive the quasi-particle excitations by 
approximating the second equation~\eqref{evolution} via 
$i\partial_t \hat\rho_{\mu\nu}^{\rm corr}=
F_2(\hat\rho_\mu^0,\hat\rho_{\mu\nu}^{\rm corr},0)+\ord(1/Z^2)$
which yields a linear equation for $\hat\rho_{\mu\nu}^{\rm corr}$. 
To solve this linear equation, it is useful to split the original 
annihilation operator
\bea
\label{full+empty} 
\hat c_{\mu\uparrow}
=
\ket{\downarrow}_\mu\!\bra{\uparrow\downarrow}+\ket{0}_\mu\!\bra{\uparrow}
=\hat c_{\mu\uparrow}\hat n_\mu^\downarrow+
\hat c_{\mu\uparrow}(1-\hat n_\mu^\downarrow) 
=
\hat f_{\mu\uparrow}+\hat e_{\mu\uparrow}^\dagger
\ea
into the annihilation operator $\hat f_{\mu\uparrow}$ of a full lattice site 
$\ket{\uparrow\downarrow}_\mu$ and the creation operator 
$\hat e_{\mu\uparrow}^\dagger$ of an empty lattice site $\ket{0}_\mu$. 
After a spatial Fourier transform (assuming infinite-size lattices), 
the linear equation for 
$\hat\rho_{\mu\nu}^{\rm corr}$ can be mapped onto a set of linear equations 
for the operators $\hat f_{\mathbf{k}s}$ and $\hat e_{\mathbf{k}s}^\dagger$
\bea
\label{matrix}
i\partial_t
\left(
\begin{array}{c}
\hat f_{\mathbf{k}s}
\\
\hat e_{\mathbf{k}s}^\dagger 
\end{array}
\right)
=
\left(
\begin{array}{cc}
U-T_{\bf k}/2 & -T_{\bf k}/2
\\
-T_{\bf k}/2 & -T_{\bf k}/2
\end{array}
\right)
\cdot
\left(
\begin{array}{c}
\hat f_{\mathbf{k}s}
\\
\hat e_{\mathbf{k}s}^\dagger 
\end{array}
\right)
\,.
\ea
For convenience, we have re-scaled all length scales with respect to the 
lattice spacing $\ell$ and thus the wave-numbers used here are dimensionless.
Note that the wave-numbers $\mathbf{k}$ are also vectors, but 
-- depending on the dimensionality of the lattice -- possibly in a 
vector space of different dimension than the three-dimensional vector 
$\hat{\mathbf{S}}$ in Eq.~\eqref{spin}.

Diagonalizing the above $2\times2$-matrix, the eigenvalues 
$\lambda^\pm_\mathbf{k}$ yield the quasi-particle energies 
$E^\pm_\mathbf{k}$ via 
\bea
\label{quasi-particle-energies}
\lambda^\pm_\mathbf{k}
=
\pm E^\pm_\mathbf{k}
=
\frac12\left(U-T_\mathbf{k}\pm\sqrt{T_\mathbf{k}^2+U^2}\right)
\,,
\ea
where $T_\mathbf{k}$ denotes the Fourier transform of the hopping matrix 
$T_{\mu\nu}$. 
In the strong-coupling limit $U\gg T$, these quasi-particle energies simplify 
to $E^+_\mathbf{k}\approx U-T_\mathbf{k}/2$ and 
$E^-_\mathbf{k}\approx T_\mathbf{k}/2$.   

Starting from the grand-canonical Hamiltonian including the chemical
potential, the matrix in Eq.~\eqref{matrix} would contain $\pm U/2$ on the 
diagonal and thus its eigenvalues would be lowered by $U/2$ such that the 
quasi-particle energies in Eq.~\eqref{quasi-particle-energies} 
assume a more symmetric form 
$E^\pm_\mathbf{k}\approx U/2\mp T_\mathbf{k}/2$.
In the following, we shall use the convention~\eqref{quasi-particle-energies}.

Note that, starting from the mean-field ansatz~\eqref{Ising}
reflecting the perfect Ising type anti-ferromagnetic order, 
the quasi-particle energies would not 
contain such a contribution linear in $T_\mathbf{k}$ but scale quadratically 
$\ord(T_\mathbf{k}^2/U)$. 
As an intuitive picture, since neighboring sites always have opposite spins, 
the propagation of doublons or holons on such a perfectly spin-ordered 
background can only occur via second-order tunneling processes. 
In contrast, for Heisenberg type spin order (or an unordered state), 
there is a finite probability that neighboring lattice sites are occupied 
by particles with the same spin, such that doublons or holons can propagate 
via first-order tunneling processes. 


The eigenvectors of the matrix in Eq.~\eqref{matrix} determine the Bogoliubov 
transformation to the quasi-particle operators $\hat d_{\mathbf{k}s}$ and 
$\hat h_{\mathbf{k}s}^\dagger$
\bea
\label{Bogoliubov}
\left(
\begin{array}{c}
\hat d_{\mathbf{k}s}
\\
\hat h_{\mathbf{k}s}^\dagger 
\end{array}
\right)
=
\left(
\begin{array}{cc}
\cos\varphi_\mathbf{k} & \sin\varphi_\mathbf{k}
\\
-\sin\varphi_\mathbf{k} & \cos\varphi_\mathbf{k}
\end{array}
\right)
\cdot
\left(
\begin{array}{c}
\hat f_{\mathbf{k}s}
\\
\hat e_{\mathbf{k}s}^\dagger 
\end{array}
\right)
\,,
\ea
with the rotation angle 
\bea 
\tan\varphi_\mathbf{k}=\frac{\sqrt{T_\mathbf{k}^2+U^2}+U}{T_\mathbf{k}}
\,,
\ea
where $\hat d_{\mathbf{k}s}$ is the annihilation operator for a doublon 
with energy $E^+_\mathbf{k}$ while $\hat h_{\mathbf{k}s}^\dagger$ is the 
creation operator of a holon with energy $E^-_\mathbf{k}$. 

Of course, the above derivation of the quasi-particle picture is not unique, 
one can also derive it via other means, e.g., the Hubbard approximation \cite{Hub63,Herr97}. 
%
%
However, the $1/Z$-expansion provides a clear and controlled path to 
incorporate higher orders consistently.

\subsection{Boltzmann equation}  

To first order in $1/Z$, the time evolution of the operators 
$\hat d_{\mathbf{k}s}$ and $\hat h_{\mathbf{k}s}$ is simply governed by 
the trivial phase factors $\exp\{-iE^\pm_\mathbf{k}t\}$ 
such that their populations 
$\langle\hat d_{\mathbf{k}s}^\dagger\hat d_{\mathbf{k}s}\rangle$ and 
$\langle\hat h_{\mathbf{k}s}^\dagger\hat h_{\mathbf{k}s}\rangle$ remain 
constant. 
Interactions such as collisions between these quasi-particles leading to 
a finite energy and momentum transfer would induce a redistribution of these 
populations and are thus not described within this first-order approach.
To incorporate such interactions, one has to include higher orders in $1/Z$.

To second order $1/Z^2$, one should take the three-point correlator 
$\hat\rho_{\mu\nu\lambda}^{\rm corr}$ in the second equation~\eqref{evolution} 
into account. 
Its time-derivative does also contain the four-point correlator 
$\hat\rho_{\mu\nu\lambda\kappa}^{\rm corr}$, which is of order $1/Z^3$. 
Truncating the set of evolution equations~\eqref{evolution} at this order,
i.e., neglecting all terms scaling with $1/Z^4$ or higher, 
we may apply basically the same steps (Markov approximation etc.) 
as for weakly interacting systems and arrive at a Boltzmann equation 
describing the redistribution of the quasi-particle populations
$\mathfrak{d}_\mathbf{k}^s=
\langle\hat d_{\mathbf{k}s}^\dagger\hat d_{\mathbf{k}s}\rangle$ 
and 
$\mathfrak{h}_\mathbf{k}^s=
\langle\hat h_{\mathbf{k}s}^\dagger\hat h_{\mathbf{k}s}\rangle$. 
Focusing on the holon sector for simplicity, we find in the strong-coupling
limit $U\gg T$ (where $E^-_\mathbf{k}\approx T_\mathbf{k}/2$)  
for the mean-field background~\eqref{mean-field},
see App.~\ref{Derivation of Boltzmann equation} 
and \ref{Doublons and holons} \cite{Queiss19b,Queiss19c}
\bea
\label{Boltzmann}
\partial_t \mathfrak{h}_\mathbf{k}^\uparrow
&=& 
-2\pi \int\limits_{\mathbf{p}\mathbf{q}}
\left(T_\mathbf{k}+T_\mathbf{p}\right)^2 
\delta\left(
E_\mathbf{k}^-
+E_\mathbf{p}^-
-E_\mathbf{k+q}^-
-E_\mathbf{p-q}^-
\right)
\nn
&&
\times 
\left[
\mathfrak{h}_\mathbf{k}^\uparrow
\mathfrak{h}_\mathbf{p}^\downarrow
\left(1-\mathfrak{h}_\mathbf{k+q}^\uparrow\right)
\left(1-\mathfrak{h}_\mathbf{p-q}^\downarrow\right)
-
\mathfrak{h}_\mathbf{k+q}^\uparrow
\mathfrak{h}_\mathbf{p-q}^\downarrow
\left(1-\mathfrak{h}_\mathbf{k}^\uparrow\right)
\left(1-\mathfrak{h}_\mathbf{p}^\downarrow\right)
\right] 
\,.
\ea
Thus, even in the strongly interacting limit, the quasi-particle 
distributions obey a Boltzmann equation which has the usual interpretation: 
Two holons with opposite spins and  initial momenta $\mathbf{k}$ and 
$\mathbf{p}$ collide with each other and are scattered to the final momenta 
$\mathbf{k+q}$ and $\mathbf{p-q}$ where $\mathbf{q}$ is the momentum transfer. 
Note that the scattering cross section $\propto(T_\mathbf{k}+T_\mathbf{p})^2$
is actually independent of the momentum transfer $\mathbf{q}$. 
For two holons with the same spin, we found a vanishing scattering cross 
section, i.e., they do not interact at this order.
%

The term in the third line of Eq.~\eqref{Boltzmann} represents the inverse 
process and ensures the conservation of probability or total holon number. 
Energy conservation is implied by the Dirac delta distribution in the first 
line of Eq.~\eqref{Boltzmann}.
Since the above Boltzmann equation~\eqref{Boltzmann} assumes the standard 
form, it entails the usual consequences, such as the $H$-theorem describing 
thermalization etc. 
%

Focusing on the doublon sector instead, one obtains precisely the same form 
of the Boltzmann equation~\eqref{Boltzmann} for 
$\mathfrak{d}_\mathbf{k}^s=
\langle\hat d_{\mathbf{k}s}^\dagger\hat d_{\mathbf{k}s}\rangle$ 
instead of 
$\mathfrak{h}_\mathbf{k}^s=
\langle\hat h_{\mathbf{k}s}^\dagger\hat h_{\mathbf{k}s}\rangle$, 
as expected from the particle-hole duality mentioned in the Introduction. 
Taking both sectors into account simultaneously also accounts for 
collisions between doublons and holons, see 
Appendices~\ref{Derivation of Boltzmann equation} 
and \ref{Doublons and holons}. 

Note that initial states which are spin polarized in $\sigma_x$ 
direction, for example, would also induce off-diagonal terms such as 
$\langle\hat h_{\mathbf{k}\uparrow}^\dagger
\hat h_{\mathbf{k}\downarrow}\rangle$, see also \cite{Ba22}.
In the absence of such a spin polarization, however, these terms vanish 
initially and thus stay zero throughout the evolution because our 
equations of motion do not contain symmetry-breaking contributions 
such as magnetic fields. 
Thus, we omit these off-diagonal terms here. 

\section{Effective Hamiltonian}\label{Effective Hamiltonian}

In order to compare the Boltzmann equation~\eqref{Boltzmann} obtained via the
$1/Z$-expansion with the standard derivation of Boltzmann equations for weakly 
interacting systems, let us construct an effective Hamiltonian which would 
reproduce Eq.~\eqref{Boltzmann} in this way.
To this end, let us start with the usual fermionic creation and annihilation
operators $\hat a_{\mathbf{k}s}^\dagger$ and $\hat a_{\mathbf{k}s}$ and 
the standard ansatz for such an effective Hamiltonian
%
%
\bea
\label{effective}
\hat H_{\rm eff}
= 
\sum_s\int\limits_{\mathbf{k}}
E_\mathbf{k} \hat a_{\mathbf{k}s}^\dagger \hat a_{\mathbf{k}s}
+
\int\limits_{\mathbf{k}\mathbf{p}\mathbf{q}}
V_{\mathbf{k}\mathbf{p}\mathbf{q}}^{\uparrow\downarrow}
\hat a_{\mathbf{k+q}\uparrow}^\dagger 
\hat a_{\mathbf{p-q}\downarrow}^\dagger  
\hat a_{\mathbf{p}\downarrow} 
\hat a_{\mathbf{k}\uparrow} 
\,.
\ea
If we now set 
$V_{\mathbf{k}\mathbf{p}\mathbf{q}}^{\uparrow\downarrow}=
-(T_\mathbf{k}+T_\mathbf{p}+T_\mathbf{k+q}+T_\mathbf{p-q})/2$
as well as $E_\mathbf{k}=T_\mathbf{k}/2$, 
we would indeed recover Eq.~\eqref{Boltzmann} via the usual 
Born-Markov approximation.

However, a few cautionary remarks are in order. 
First, the standard derivation of Eq.~\eqref{Boltzmann} from 
Eq.~\eqref{effective} is based on the usual fermionic commutation 
relations between the operators $\hat a_{\mathbf{k}s}^\dagger$ and 
$\hat a_{\mathbf{k}s}$.
In contrast, neither the doublon 
$\hat d_{\mathbf{k}s}^\dagger$ and $\hat d_{\mathbf{k}s}$ 
nor the holon operators 
$\hat h_{\mathbf{k}s}^\dagger$ and $\hat h_{\mathbf{k}s}$
satisfy these commutation relations, see Eqs.~\eqref{full+empty} and 
\eqref{Bogoliubov}. 
Second, in contrast to the weakly interacting case, both 
$V_{\mathbf{k}\mathbf{p}\mathbf{q}}^{\uparrow\downarrow}$ 
and $E_\mathbf{k}$ scale with 
$T_\mathbf{k}$ and are thus not really independent, which requires 
special care when justifying the Born-Markov approximation.
%
%
%
It should also be noted here that the insertion of the simple replacement 
$\hat a_{\mu\uparrow}\to\hat a_{\mu\uparrow}
(1-\hat a_{\mu\downarrow}^\dagger\hat a_{\mu\downarrow})$
into the free Hamiltonian 
$\sum_{\mu\nu s} T_{\mu\nu} \hat a_{\mu s}^\dagger \hat a_{\nu s}$  
does not yield the correct effective Hamiltonian~\eqref{effective}.

As another point, the scattering cross section in the Boltzmann 
equation~\eqref{Boltzmann} is given by the square of the interaction 
matrix element $|V^{\uparrow\downarrow}_{\mathbf{k}\mathbf{p}\mathbf{q}}|^2$ 
and thus 
does not uniquely determine the sign (or phase) of 
$V^{\uparrow\downarrow}_{\mathbf{k}\mathbf{p}\mathbf{q}}$, 
e.g., whether the interaction 
is attractive or repulsive.
For example, for doublons one should insert 
$E_\mathbf{k}=U-T_\mathbf{k}/2$
and $V_{\mathbf{k}\mathbf{p}\mathbf{q}}^{\uparrow\downarrow}=
(T_\mathbf{k}+T_\mathbf{p}+T_\mathbf{k+q}+T_\mathbf{p-q})/2$ 
into the effective Hamiltonian~\eqref{effective}, which does, however, 
yield the same Boltzmann equation~\eqref{Boltzmann}. 

\section{Perturbation Theory in $T/U$}\label{Perturbation Theory}

In order to settle the sign ambiguity 
mentioned above, let us compare our results 
to strong-coupling perturbation theory, i.e., a power expansion in the small 
control parameter $\epsilon=T/U\ll1$. 
To this end, we split the Hamiltonian~\eqref{Fermi-Hubbard} via 
$\hat H=\hat H_U+\hat H_T=\hat H_0+\hat H_1$ into an undisturbed part 
$\hat H_0=\hat H_U=\ord(\epsilon^0)$ plus a perturbation 
$\hat H_1=\hat H_T=\ord(\epsilon^1)$. 
For general matrix elements 
\bea
{\cal M}=\bra{\Psi_{\rm out}}\hat H\ket{\Psi_{\rm in}}\,,
\ea
we employ the same power expansion of the states 
\bea
\ket{\Psi_{\rm in}}
=
\ket{\Psi_{\rm in}}_0+\epsilon\ket{\Psi_{\rm in}}_1+\ord(\epsilon^2)
\,,
\ea
and analogously for $\ket{\Psi_{\rm out}}$. 

Because all the states considered in this section satisfy  
$\hat H_0\ket{\Psi_{\rm in}}_0=0$ and $\hat H_0\ket{\Psi_{\rm out}}_0=0$,  
the first-order matrix elements simplify to 
\bea
\label{first-order-matrix-elements}
{\cal M}=\bra{\Psi_{\rm out}}\hat H_1\ket{\Psi_{\rm in}}_0+\ord(\epsilon^2)
\,.
\ea
Apart from the power expansion in $\epsilon$, we have not made any 
assumptions regarding the states $\ket{\Psi_{\rm in}}$ and 
$\ket{\Psi_{\rm out}}$, e.g., regarding their degeneracy. 
They could be the same states, where ${\cal M}$ would yield the 
energy expectation value, or they could be different states, where 
${\cal M}$ would describe a transition matrix element. 

\subsection{Mott state} 

Let us start with the Mott state $\ket{{\rm Mott}}$, which we take to be the 
ground state of the Fermi-Hubbard Hamiltonian~\eqref{Fermi-Hubbard} at half 
filling (but other choices would also be possible).
In the quasi-particle picture, it describes the state without doublons 
$\hat d_{\mathbf{k}s}\ket{{\rm Mott}}=0$ and holons 
$\hat h_{\mathbf{k}s}\ket{{\rm Mott}}=0$. 
After a power expansion in $\epsilon$
\bea
\ket{{\rm Mott}}=\ket{{\rm Mott}}_0+\epsilon\ket{{\rm Mott}}_1+\ord(\epsilon^2)
\,,
\ea
the zeroth order $\ket{{\rm Mott}}_0$ has exactly one particle per site, i.e.,  
$\hat e_{\mathbf{k}s}\ket{{\rm Mott}}_0=0$ and 
$\hat f_{\mathbf{k}s}\ket{{\rm Mott}}_0=0$.

The virtual hopping corrections mentioned in the Introduction are included in 
the first-order correction 
\bea
\label{first-order-correction}
\ket{{\rm Mott}}_1=-\frac{\hat H_1}{U}\ket{{\rm Mott}}_0
\,,
\ea
consistent with Bogoliubov transformation~\eqref{Bogoliubov} between 
$\hat d_{\mathbf{k}s}$ and $\hat h_{\mathbf{k}s}$
on the one hand and 
$\hat f_{\mathbf{k}s}$ and $\hat e_{\mathbf{k}s}$
on the other hand. 
Obviously, the first-order energy shift vanishes 
\bea
\bra{{\rm Mott}}\hat H_1\ket{{\rm Mott}}_0=0
\,,
\ea
such that the ground-state energy is of order $T^2/U$.  

\subsection{One-holon state} 

The quasi-particle picture described above motivates the ansatz 
$\hat h_{\mathbf{k}s}^\dagger\ket{{\rm Mott}}$ for the state containing 
one holon. 
However, one should be a bit careful because the operators 
$\hat h_{\mathbf{k}s}$ and $\hat h_{\mathbf{k}s}^\dagger$ do not obey the 
usual commutation relations. 
Fortunately, the calculation of the first-order matrix 
elements~\eqref{first-order-matrix-elements} only requires the 
zeroth-order states
\bea
\ket{\Psi_{\rm in}}_0 
=
{\cal N}_{\mathbf{k}\uparrow}\hat e^\dagger_{\mathbf{k}\uparrow}\ket{{\rm Mott}}_0
=
{\cal N}_{\mathbf{k}\uparrow}\hat c_{\mathbf{k}\uparrow}\ket{{\rm Mott}}_0
=
{\cal N}_{\mathbf{k}\uparrow}\sum\limits_\alpha
\hat c_{\alpha\uparrow}\ket{{\rm Mott}}_0
\exp\{i\mathbf{k}\cdot\mathbf{r}_\alpha\}
\,,
\ea
where most of these difficulties are absent because the operators 
$\hat c_{\mathbf{k}s}$ and $\hat c_{\mathbf{k}s}^\dagger$ do satisfy the 
standard commutation relations. 
The normalization ${\cal N}_{\mathbf{k}\uparrow}$ can be derived from 
$\bra{{\rm Mott}}
\hat c^\dagger_{\alpha\uparrow}\hat c_{\beta\uparrow}
\ket{{\rm Mott}}_0
=\delta_{\alpha\beta}
\bra{{\rm Mott}}\hat n_{\alpha\uparrow}\ket{{\rm Mott}}_0$
and is -- independently of $\mathbf{k}$ -- just determined 
by the total number of particles with spin $\uparrow$. 

In analogy, we use the same ansatz for $\ket{\Psi_{\rm out}}_0$ with 
$\mathbf{k}'$ and same spin $\uparrow$ (all other matrix elements vanish)
\bea
{\cal M}
&=&
-\frac{\left|{\cal N}_\uparrow\right|^2}{Z}
\sum\limits_{\alpha\beta\mu\nu s} 
T_{\mu\nu}
\exp\{i\mathbf{k}\cdot\mathbf{r}_\alpha-i\mathbf{k}'\cdot\mathbf{r}_\beta\}
\nn
&&
\times 
\bra{{\rm Mott}}
\hat c^\dagger_{\beta\uparrow}
\hat c^\dagger_{\mu s}
\hat c_{\nu s}
\hat c_{\alpha\uparrow}
\ket{{\rm Mott}}_0
+\ord(\epsilon^2)
\,.
\ea
Since the hopping matrix $T_{\mu\nu}$ is only non-zero for $\mu\neq\nu$ 
and the state $\ket{{\rm Mott}}_0$ has exactly one particle per site, 
we may set $\alpha=\mu$ and $\beta=\nu$ or vice versa in the sum 
\bea
{\cal M}
&=&
\frac{\left|{\cal N}_\uparrow\right|^2}{Z}
\sum\limits_{\mu\nu} 
T_{\mu\nu}
\exp\{i\mathbf{k}\cdot\mathbf{r}_\mu-i\mathbf{k}'\cdot\mathbf{r}_\nu\}
\nn
&&
\times 
\big(
\bra{{\rm Mott}}
\hat n_{\mu\uparrow}
\hat n_{\nu\uparrow}
\ket{{\rm Mott}}_0
-\bra{{\rm Mott}}
\hat c^\dagger_{\mu\downarrow}
\hat c_{\mu\uparrow}
\hat c^\dagger_{\nu\uparrow}
\hat c_{\nu\downarrow}
\ket{{\rm Mott}}_0
\big)
+\ord(\epsilon^2)
\,.
\ea
%
In addition to the number correlator in the second line, we obtain the 
spin-flip term $\langle\hat S^-_\mu\hat S^+_\nu\rangle_0$ in the 
third line.

If the lattice and the state $\ket{{\rm Mott}}_0$ obey translational 
invariance, the expectation values only depend on the relative coordinate
$\mathbf{r}_\mu-\mathbf{r}_\nu$ and thus the sum over the center-of-mass 
coordinate $\mathbf{r}_\mu+\mathbf{r}_\nu$ corresponds to momentum 
conservation $\delta_{\mathbf{k}\mathbf{k}'}$.
In case of rotational invariance, the expectation values yield the same 
result for all pairs of neighbors $\mu$ and $\nu$ and thus the remaining 
sum over $\mathbf{r}_\mu-\mathbf{r}_\nu$ just yields the Fourier transform
$T_\mathbf{k}$ of the hopping matrix, i.e., 
${\cal M}\propto\delta_{\mathbf{k}\mathbf{k}'}T_\mathbf{k}+\ord(\epsilon^2)$.
For the mean-field ansatz~\eqref{mean-field}, we find 
${\cal M}=\delta_{\mathbf{k}\mathbf{k}'}T_\mathbf{k}/2+\ord(\epsilon^2)$ 
which reproduces the holon energy~\eqref{quasi-particle-energies} to lowest 
order for $\mathbf{k}=\mathbf{k}'$ and vanishes for 
$\mathbf{k}\neq\mathbf{k}'$, reflecting momentum conservation. 
As an outlook, one could study the scattering of holons 
(i.e., $\mathbf{k}\neq\mathbf{k}'$) by spin inhomogeneities via 
inserting a mean-field ansatz which breaks translational invariance. 

If we replace the mean-field ansatz~\eqref{mean-field} by the Ising 
type anti-ferromagnet~\eqref{Ising}, we find that the first-order 
matrix elements vanish ${\cal M}=\ord(\epsilon^2)$. 
Again, this is consistent with the quasi-particle picture because the 
quasi-particle energies do not contain a linear contribution in this case, 
as discussed after Eq.~\eqref{quasi-particle-energies}.

\subsection{Two-holon state} 
\label{twoholonstate}

Now let us consider initial $\ket{\Psi_{\rm in}}_0$ and final 
$\ket{\Psi_{\rm out}}_0$ states containing two holons, where we
start with the case of opposite spins, as motivated by the Boltzmann
equation~\eqref{Boltzmann}.
As usual in scattering theory, we envisage initial and final holon 
wave-packets which do not overlap but interact in an intermediate 
space-time region. 
Then, in straightforward generalization of the one-holon case, we use the 
following ansatz for their Fourier components 
\bea
\label{two-holon-ansatz}
\ket{\Psi_{\rm in}}_0 
=
{\cal N}_{\mathbf{k}_1\mathbf{k}_2}^{\uparrow\downarrow}
\hat c_{\mathbf{k}_1\uparrow}\hat c_{\mathbf{k}_2\downarrow}
\ket{{\rm Mott}}_0
=
{\cal N}_{\mathbf{k}_1\mathbf{k}_2}^{\uparrow\downarrow}
\sum\limits_{\alpha\beta} 
\hat c_{\alpha\uparrow}\hat c_{\beta\downarrow}
\ket{{\rm Mott}}_0
e^{i\mathbf{k}_1\cdot\mathbf{r}_\alpha+i\mathbf{k}_2\cdot\mathbf{r}_\beta}
\,,
\ea
and analogously for $\ket{\Psi_{\rm out}}_0$ with $\mathbf{k}_3$ and 
$\mathbf{k}_4$.  
The resulting matrix elements read 
\bea
{\cal M}
&=&
-
\frac{
{\cal N}_{\mathbf{k}_1\mathbf{k}_2}^{\uparrow\downarrow}
({\cal N}_{\mathbf{k}_3\mathbf{k}_4}^{\uparrow\downarrow})^*
}{Z}
\sum\limits_{\alpha\beta\gamma\delta\mu\nu s}
e^{i\mathbf{k}_1\cdot\mathbf{r}_\alpha+i\mathbf{k}_2\cdot\mathbf{r}_\beta
-i\mathbf{k}_3\cdot\mathbf{r}_\gamma-i\mathbf{k}_4\cdot\mathbf{r}_\delta}
\nn
&&
\times 
T_{\mu\nu}
\bra{{\rm Mott}}
\hat c_{\delta\downarrow}^\dagger 
\hat c_{\gamma\uparrow}^\dagger 
\hat c^\dagger_{\mu s}
\hat c_{\nu s}
\hat c_{\alpha\uparrow}\hat c_{\beta\downarrow}
\ket{{\rm Mott}}_0
+\ord(\epsilon^2)
\,.
\ea
The expectation values in the second line 
%
%
%
are only non-zero if $\alpha$, $\beta$, and $\nu$ are mutually different, 
and the same for $\gamma$, $\delta$, and $\mu$.
We only get non-vanishing contributions if the triple $\{\alpha,\beta,\nu\}$ 
is a permutation of the triple $\{\gamma,\delta,\mu\}$.
In view of $T_{\mu\nu}=0$ for $\mu=\nu$, we are left with four permutations 
(and the sum over spin $s$). 

Altogether, this yields expectation values of the number operators such as 
$\bra{{\rm Mott}}
\hat n_\alpha^\uparrow\hat n_\mu^\downarrow\hat n_\nu^\downarrow
\ket{{\rm Mott}}_0$
and spin-flip terms of the form 
$\bra{{\rm Mott}}\hat n_\alpha^\uparrow\hat S_\mu^+\hat S_\nu^-\ket{{\rm Mott}}_0$.
%
%
For the mean-field ansatz~\eqref{mean-field}, only the former contribute and the 
matrix element simplifies to 
\bea
\label{matrix-element}
{\cal M}
&=&
\frac{T_{\mathbf{k}_1}+T_{\mathbf{k}_2}}{2}\, 
\delta_{\mathbf{k}_1\mathbf{k}_3}\delta_{\mathbf{k}_2\mathbf{k}_4}
-\frac{T_{\mathbf{k}_1}+T_{\mathbf{k}_2}+T_{\mathbf{k}_3}+T_{\mathbf{k}_4}}{2}\,
\delta_{\mathbf{k}_1+\mathbf{k}_2,\mathbf{k}_3+\mathbf{k}_4}
\,.
\ea
This result is consistent with the effective Hamiltonian~\eqref{effective}
where the first term on the right-hand side of Eq.~\eqref{matrix-element} 
corresponds to the free propagation of the two holons with 
their quasi-particle energies $E_\mathbf{k}$ while the second term
describes their scattering with the effective interaction
potential $V^{\uparrow\downarrow}_{\mathbf{k}\mathbf{p}\mathbf{q}}$. 

The origin of this effective interaction potential is the fact that the sums 
are not independent of each other, e.g., the sum over $\alpha=\gamma$ is not 
independent of the remaining sums over $\mu=\beta$ and $\nu=\delta$ 
because $\alpha$, $\beta$, and $\nu$ must be mutually different to yield 
a non-zero expectation value (as explained above).
As an intuitive picture, the presence of the $\uparrow$-holon at site $\alpha$
may effectively inhibit
the hopping of the $\downarrow$-holon from site $\nu$ to 
$\mu$ and thus changes its energy -- which implies an effective interaction. 


\subsection{Two-holon triplet state} 

For comparison, let us consider the state of two holons with the same spin. 
In complete analogy to Eq.~\eqref{two-holon-ansatz}, we use the ansatz 
\bea
\label{two-holon-triplet}
\ket{\Psi_{\rm in}}_0 
&=&
{\cal N}_{\mathbf{k}_1\mathbf{k}_2}^{\uparrow\uparrow}
\hat c_{\mathbf{k}_1\uparrow}\hat c_{\mathbf{k}_2\uparrow}
\ket{{\rm Mott}}_0
\,.
\ea
Following the same steps as in the previous subsection, 
including the insertion of the mean-field ansatz~\eqref{mean-field}, 
we find that only the matrix elements corresponding to the 
free propagation survive 
\bea
{\cal M}
=
\frac{T_{\mathbf{k}_1}+T_{\mathbf{k}_2}}{2}
\left(
\delta_{\mathbf{k}_1\mathbf{k}_3}\delta_{\mathbf{k}_2\mathbf{k}_4}
-
\delta_{\mathbf{k}_1\mathbf{k}_4}\delta_{\mathbf{k}_2\mathbf{k}_3}
\right) 
\,.
\ea
Again, this is consistent with the Boltzmann equation~\eqref{Boltzmann}
which also did not contain scattering between holons of equal spin. 

\subsection{Spin correlations}  

So far, our results were based on the zeroth-order mean-field 
ansatz~\eqref{mean-field} which neglects all correlations between 
the lattice sites. 
Including such correlations leads to corrections to these results. 
For the two-holon triplet state, for example, an effective interaction 
$V_{\mathbf{k}\mathbf{p}\mathbf{q}}^{\uparrow\uparrow}$
can be obtained if we include correlations between lattice sites, 
i.e., go beyond the mean-field ansatz~\eqref{mean-field}. 
Taking into account these correlations between two lattice sites 
$\mu$ and $\nu$ as encoded in $\hat\rho_{\mu\nu}^{\rm corr}=\ord(1/Z)$, 
but neglecting all three-point correlators 
$\hat\rho_{\mu\nu\lambda}^{\rm corr}=\ord(1/Z^2)$, we find  
\bea
\label{same-spin}
V_{\mathbf{k}\mathbf{p}\mathbf{q}}^{\uparrow\uparrow}
=
\left(
T_{\mathbf{k}}+T_{\mathbf{p}}+T_{\mathbf{k+q}}+T_{\mathbf{p-q}}
\right) 
C_\mathbf{q}^{\uparrow\uparrow}
\,,
\ea
where $C_\mathbf{q}^{\uparrow\uparrow}$ denotes the Fourier transform of the 
number correlations 
\bea
\bra{{\rm Mott}}\hat n_\alpha^\uparrow\hat n_\beta^\uparrow
\ket{{\rm Mott}}_0^{\rm corr}
=
\int\limits_\mathbf{q}C_\mathbf{q}^{\uparrow\uparrow}
e^{i\mathbf{q}\cdot(\mathbf{r}_\alpha-\mathbf{r}_\beta)}
\,.
\ea
The sign of the correlations depends on the spin order of the background state.
For anti-ferromagnetic order, 
$\langle\hat n_\alpha^\uparrow\hat n_\beta^\uparrow\rangle_0^{\rm corr}$
is negative for nearest neighbors $\alpha$ and $\beta$ but positive for 
next-to-nearest neighbors -- while for (locally) ferromagnetic order, 
it would also be positive for nearest neighbors.
%

In analogy, we may derive the correlation corrections to the interaction 
between two holons of opposite spin 
\bea
\label{opposite-spin}
V_{\mathbf{k}\mathbf{p}\mathbf{q}}^{\uparrow\downarrow}
&=&
-\frac12
\left(
T_{\mathbf{k}}+T_{\mathbf{p}}+T_{\mathbf{k+q}}+T_{\mathbf{p-q}}
\right) 
\left[1-4C_\mathbf{q}^{\uparrow\downarrow}-
8C_\mathbf{p-k-q}^{\uparrow\downarrow}\right]
\,,
\ea
where we have used the $SU(2)$-symmetry \cite{Lieb89} 
of the Mott state
$\langle\hat S^x_\mu\hat S^x_\nu\rangle_0=
\langle\hat S^y_\mu\hat S^y_\nu\rangle_0=
\langle\hat S^z_\mu\hat S^z_\nu\rangle_0$
in order to express 
$\langle\hat S^-_\mu\hat S^+_\nu+\hat S^-_\nu\hat S^+_\mu\rangle_0
=2\langle\hat S^x_\mu\hat S^x_\nu+\hat S^y_\mu\hat S^y_\nu\rangle_0$
in terms of 
$\langle\hat S^z_\mu\hat S^z_\nu\rangle_0$, i.e., the number correlations
such as $\langle\hat n_\mu^\uparrow\hat n_\nu^\uparrow\rangle_0^{\rm corr}$
or 
$\langle\hat n_\mu^\uparrow\hat n_\nu^\downarrow\rangle_0^{\rm corr}=
-\langle\hat n_\mu^\uparrow\hat n_\nu^\uparrow\rangle_0^{\rm corr}$. 
%

\section{Hubbard tetramer} 
\label{Hubbard tetramer}

Let us exemplify the above results for an analytically solvable example, 
the Hubbard tetramer consisting of four lattice sites in the form of a square ($Z=2$) 
\cite{Sch01,Sch22}.
Already in this simple case, the total Hilbert space has $4^4=256$ dimensions 
and thus the Hamiltonian~\eqref{Fermi-Hubbard} can be represented by a 
$256\times256$-matrix. 
However, by using the conserved quantities such as the particle numbers
$N^\uparrow$ and $N^\downarrow$, the total spin~\eqref{spin} and 
pseudo-spin (see Appendix~\ref{APP:pseudospin}), 
as well as the spatial symmetries, 
one may cast this Hamiltonian into a block-diagonal form consisting 
of matrices with maximum rank four -- admitting analytic solutions. 

In contrast to the previous sections, which were devoted to the 
case of half filling (only marginally disturbed by one or two holons),
we shall now also consider filling factors of $3/8$ (one holon) 
and $1/4$ (two holons).

\subsection{Mott state} 

In the sector $N^\uparrow=N^\downarrow=2$, the ground state has an energy of 
$-3T^2/U+\ord(\epsilon^3)$ and vanishing total spin $S=0$ as well as 
pseudo-spin $\eta=0$, see also \cite{Lieb89}. 
We identify this state with the Mott state $\ket{\rm Mott}$, which then 
displays the lowest-order structure 
\bea
\label{Mott-zero}
\ket{\rm Mott}_0
=
\frac{
\ket{\uparrow,\downarrow,\uparrow,\downarrow}+
\ket{\downarrow,\uparrow,\downarrow,\uparrow}}{\sqrt{3}}
-\frac{
\ket{\uparrow,\uparrow,\downarrow,\downarrow}+
\ket{\downarrow,\uparrow,\uparrow,\downarrow}+
\ket{\downarrow,\downarrow,\uparrow,\uparrow}+
\ket{\uparrow,\downarrow,\downarrow,\uparrow}
}{\sqrt{12}}
\,.
\nn
\ea
The first-order hopping corrections can be obtained from 
Eq.~\eqref{first-order-correction}. 
Note that one should be careful with the above representation 
because the sign of the basis vectors such as 
$\ket{\uparrow,\downarrow,\uparrow,\downarrow}=
\hat c_{4\downarrow}^\dagger\hat c_{3\uparrow}^\dagger
\hat c_{2\downarrow}^\dagger\hat c_{1\uparrow}^\dagger
\ket{0}$ 
depends on the chosen order of the fermionic operators. 
Note that there is also another state in this singlet sector 
with $S=\eta=0$ and $N^\uparrow=N^\downarrow=2$, which has 
a slightly higher energy of $-T^2/U+\ord(\epsilon^3)$. 
%

\subsection{One-holon state} 

The one-holon states -- as single quasi-particle excitations
around the Mott state -- are then identified with the eigenstates 
in the doublet sector with $S=1/2$ and 
$N^\uparrow=2$ and $N^\downarrow=1$ (or $N^\uparrow=1$ and $N^\downarrow=2$).
They have eigen-energies of $\pm T/2+\ord(\epsilon^2)$ and 
$\pm\sqrt{3}T/2+\ord(\epsilon^2)$. 
In the following, we shall omit the symbols $\ord(\epsilon^2)$ for 
brevity and just state the energies to first order in $T$. 

\subsection{Two-holon state} 

To study the two-holon states, let us first consider the case  
$N^\uparrow=N^\downarrow=1$. 
These two-holon states lie in the singlet ($S=0$) or triplet ($S=1$) 
sector and have eigen-energies $\pm\sqrt{2}T$, $\pm T$, and zero. 
The fact that all eigen-energies obey the reflection symmetry $T\to-T$
is a consequence of the staggered gauge transformation mentioned in the 
Introduction.

Already on the level of the eigen-energies, we find that not all 
two-holon energies can be written as a sum of two one-holon energies -- 
which can be interpreted as a signature of their interactions. 
For example, adding one $\uparrow$ holon with energy $-\sqrt{3}T/2$
to another $\downarrow$ holon with the same energy $-\sqrt{3}T/2$, 
one would expect a total energy of $-\sqrt{3}T$ 
in the non-interacting case.
However, such an energy is not contained in the spectrum. 
Instead, the lowest two-holon energy is $-\sqrt{2}T$.  
As an intuitive picture, the presence of the $\downarrow$ holon
reduces the options for the $\uparrow$ holon to lower the energy 
via tunneling and vice versa. 
As a result, these two quasi-particles effectively repel each other
in this case. 

In addition to the eigen-energies, we may also consider the eigen-states.
To this end, let us introduce the operators 
\bea
\hat c_{s\pm}
=
\frac{1}{\sqrt{2}} 
\sum_{\mu=1}^4 (\pm1)^\mu\hat c_{\mu s} 
\,.
\ea
Acting on the lowest-order Mott state~\eqref{Mott-zero}, 
these operators generate the lowest-order one-holon states 
with energies $\mp T/2$. 
However, if we generate a two-holon state by applying these operators 
twice $\hat c_{\uparrow+}\hat c_{\downarrow+}\ket{\rm Mott}_0$, 
we obtain an eigen-state with zero energy. 
This again supports the interpretation that one holon disturbs the 
hopping options for the other holon such that they repel each other. 
Note that the same zero-energy state can be obtained via 
$\hat c_{\uparrow-}\hat c_{\downarrow-}\ket{\rm Mott}_0$
which is consistent with the staggered gauge transformation 
mentioned in the Introduction and would then lead to the 
interpretation that these two holons attract each other. 
Another zero-energy eigenstate can be obtained by 
$\hat c_{\uparrow+}\hat c_{\downarrow-}\ket{\rm Mott}_0$
which would fit to the non-interacting case. 

\subsection{Two-holon triplet state} 

To complete the picture, let us discuss the two-holon states 
for the case $N^\uparrow=2$ and $N^\downarrow=0$. 
Obviously, they are in the triplet sector $S=1$ and 
the repulsion $U$ does not play any role in this case.
Thus the eigen-energies are the same as in the non-interacting case, 
i.e., $\pm T$ and zero.
If we try to write these two-holon eigen-energies as the sum of two 
one-holon eigen-energies, we see that this works for some of the 
one-holon states (with energies $\pm T/2$), but not for the others 
(with energies $\pm\sqrt{3}T/2$), which can again be interpreted 
as a signature of their interactions. 
Even though the repulsion $U$ does not play any role for the two-holon
states with $N^\uparrow=2$ and $N^\downarrow=0$, it is important for the 
one-holon states. 

As an example for the states, we can obtain a two-holon eigen-state via 
$\hat c_{\downarrow+}\hat c_{\downarrow-}\ket{\rm Mott}_0$ which has 
(exactly) zero eigen-energy. 
Consistent with the above considerations, this would correspond to a 
case where two holons do not interact. 

\section{Bardeen-Cooper-Schrieffer (BCS) Theory} 
\label{Bardeen-Cooper-Schrieffer}

Having obtained repulsive as well as attractive contributions to the 
interaction between the quasi-particles such as holons, let us now 
investigate possible implications for Bardeen-Cooper-Schrieffer (BCS) 
like pairing, which might be relevant for our understanding of 
high-temperature superconductivity \cite{Dag94,Scal95,Bed86}.
To this end, we assume a small but finite density of holons -- 
corresponding to a filling factor slightly below half filling.

As one possible approach, one could start from an effective Hamiltonian 
such as in Eq.~\eqref{effective} and then follow a procedure very analogous
to the standard BCS theory of superconductivity \cite{Ba57}, 
see Section~\ref{BCS-effective-Hamiltonian} below.
However, as already explained in Section~\ref{Effective Hamiltonian}, 
one might object that the effective quasi-particle operators do not obey 
the standard commutation relations. 

\subsection{Variational ansatz}

Thus, we shall first pursue a more conservative approach and employ a 
variational 
ansatz in order to see whether and when BCS like pairing could lead to a 
reduction of the energy. 
To this end, we use the following ansatz for the zeroth-order BCS state 
\bea 
\ket{\rm BCS}_0
=
{\cal N}\exp\left\{\sum\limits_{\mu\nu}\xi_{\mu\nu}
\hat c_{\mu\uparrow}\hat c_{\nu\downarrow}
\right\} 
\ket{\rm Mott}_0
\,,
\ea 
with the pairing (squeezing) operator $\xi_{\mu\nu}$ and a 
normalization $\cal N$ which is required because the above 
exponential is not unitary. 
At a first glance, this ansatz may appear a bit unusual, 
but using translational invariance of the $\xi_{\mu\nu}$,
we may cast it into a more familiar form 
\bea 
\ket{\rm BCS}_0
=
{\cal N}\exp\left\{\int\limits_{\bf k}\xi_{\bf k}
\hat c_{\bf k\uparrow}\hat c_{-\bf k\downarrow}
\right\} 
\ket{\rm Mott}_0
=
\prod\limits_{\bf k}
\left(u_{\bf k}+v_{\bf k}\hat c_{\bf k\uparrow}\hat c_{-\bf k\downarrow}
\right)\ket{\rm Mott}_0
\,, 
\ea 
%
%
where we have used the fact that the exponential factorizes and its 
Taylor expansion terminates after the first order due to the Pauli 
principle.

Unfortunately, the Mott state does not factorize in the $\bf k$ basis, 
rendering the calculation of expectation values difficult.
Thus, we use a Taylor expansion for small $\xi_{\bf k}$ in order to study 
in which direction the energy could be reduced. 
In addition, we employ strong-coupling perturbation theory as in 
Sec.~\ref{Perturbation Theory} which yields expectation values that 
we have already calculated there 
\bea
\label{variational}
\langle\hat H\rangle
&=& 
\bra{\rm BCS}\hat H_1\ket{\rm BCS}_0
+\ord(\epsilon^2)
\nn
&=&
2\int\limits_{\bf k}|\xi_{\bf k}|^2E_{\bf k}^-
-\int\limits_{\bf k,p}\xi_{\bf k}\xi_{\bf p}^*
\left(T_{\mathbf{k}}+T_{\mathbf{p}}\right) 
\left[1-12C_\mathbf{k+p}^{\uparrow\downarrow}\right]
+\ord(\epsilon^2)
+\ord(|\xi_{\bf k}|^4)
\,.
\ea
In the first term, $|\xi_{\bf k}|^2$ just gives the number of holon pairs 
with the holon eigen-energies $E_{\bf k}^-$ which are, 
up to small correlation induced corrections, given by $T_{\bf k}/2$. 
The second term corresponds to their interaction. 

In order to avoid disturbing the Mott background too much, we consider 
a small number of holons, which is consistent with the assumption of 
small $|\xi_{\bf k}|^2$.
Then, only states close the minimum energies 
$E_{\bf k}^-\approx T_{\bf k}/2$
should be occupied by holons. 
Assuming a square lattice where $T_{\bf k}$ behaves as $\cos k_x+\cos k_y$,
states around $(k_x,k_y)=(\pi,\pi)$ are filled up first in order to 
minimize the energy.
For these states, the lowest-order interaction term 
$(T_{\mathbf{k}}+T_{\mathbf{p}})$ is repulsive, such that the usual s-wave 
pairing mechanism would not lead to a reduced energy. 

However, for d-wave order parameters $\xi_{\bf k}$, which behave as 
$\cos k_x-\cos k_y$, both $\int_{\bf k}\xi_{\bf k}$ and 
$\int_{\bf k}\xi_{\bf k}T_{\bf k}$ vanish due to the angular average 
and thus the lowest-order repulsion term $(T_{\mathbf{k}}+T_{\mathbf{p}})$
cancels.
The remaining correlations $C_\mathbf{k+p}^{\uparrow\downarrow}$ can then 
indeed favor d-wave pairing of holons since it corresponds to an effectively 
attractive contribution. 
In order to see how such a d-wave pairing could lower the energy, let us 
Taylor expand $C_\mathbf{q}^{\uparrow\downarrow}$ for small momenta 
\bea
C_\mathbf{q}^{\uparrow\downarrow}
=
c_0+c_2{\bf q}^2+c_4(q_x^4+q_y^4) 
+\tilde c_4q_x^2q_y^2
+\dots
\ea
After insertion into the variational ansatz~\eqref{variational}, 
the constant $c_0$ and quadratic $c_2$ contributions vanish after 
their convolution with the d-wave order parameters 
$\xi_{\bf k}$ and $\xi_{\bf p}^*$, 
but the quartic term $c_4$ does indeed generate a reduction of 
the energy $\langle\hat H\rangle$ provided that it is positive 
$c_4>0$.
This condition $c_4>0$ is satisfied for anti-ferromagnetic 
nearest-neighbor correlations, 
for which $C_\mathbf{q}^{\uparrow\downarrow}$ behaves as 
$\cos q_x+\cos q_y$ such that $c_4>0$ and $\tilde c_4=0$.
On the other hand, a non-zero $\tilde c_4$ could also support 
tilted d-wave pairing where $\xi_{\bf k}$ behaves as $\sin k_x \sin k_y$. 

In summary, starting with the Mott state and adding a small amount of 
holon pairs suggests an instability towards d-wave pairing 
(but not s-wave pairing).
This is generated by the effectively attractive contribution to the 
interaction between holons stemming from the correlation 
$C_\mathbf{q}^{\uparrow\downarrow}$.

\subsection{Effective Hamiltonian}\label{BCS-effective-Hamiltonian}

Of course, it would be desirable to go beyond lowest order in $\xi_{\bf k}$ 
and to include the chemical potential $\mu$ etc.
Note that $\mu$ is now meant to describe the effective chemical potential 
associated to the finite density of holons, not the chemical potential 
$\mu=U/2$ in the grand-canonical Hamiltonian for the original fermions 
as discussed after Eq.~\eqref{spin}.
Ignoring the problems associated with the effective 
Hamiltonian~\eqref{effective} for a moment, let us treat the holons 
as fundamental particles as described by the creation and annihilation 
operators $\hat{a}_{\mathbf{k},s}^\dagger$ and $\hat{a}_{\mathbf{k},s}$ 
which obey the usual fermionic commutation relations. 
Then we may start from the effective Hamiltonian~\eqref{effective}
together with the effective interaction~\eqref{opposite-spin} and 
perform the same steps as in standard BCS theory, including the 
derivation of a gap equation. 
To this end, we make an ansatz for the BCS state which is of the standard form \cite{Ba57}
\begin{align}
|\mathrm{BCS}\rangle 
=
\prod_\mathbf{k}\left(u_\mathbf{k}+v_\mathbf{k}
\hat{a}^\dagger_{-\mathbf{k},\uparrow}\hat{a}_{\mathbf{k},\downarrow}^\dagger\right)|0\rangle\,. 
\end{align}
Here $|0\rangle$ denotes the vacuum state 
$\hat{a}_{\mathbf{k},s}|0\rangle=0$ and the variational coefficients 
fulfil $|u_\mathbf{k}|^2+|v_\mathbf{k}|^2=1$ which guarantees 
the normalization of the BCS state.

The minimization of the energy 
$\langle\mathrm{BCS}|\hat{H}_\mathrm{eff}|\mathrm{BCS}\rangle$ 
leads to the self-consistency equation for the pairing amplitude 
%
\begin{align}\label{gapequ}
\triangle_\mathbf{p}
=
-\int\limits_\mathbf{k}V^{\uparrow\downarrow}_{\mathbf{k,-k,p-k}}
\langle \mathrm{BCS}|\hat{a}_{\mathbf{k},\downarrow}
\hat{a}_{-\mathbf{k},\uparrow}|\mathrm{BCS}\rangle
=
-\int\limits_\mathbf{k}V^{\uparrow\downarrow}_\mathbf{k,-k,p-k} \frac{ \triangle_\mathbf{k}}{2\sqrt{(E_\mathbf{k}-\mu)^2+\triangle_\mathbf{k}^2}}
\,.
\end{align}
As usual, non-trivial solutions $\triangle_\mathbf{p}$ are obtained if the 
interaction $V^{\uparrow\downarrow}_{\mathbf{k,-k,p-k}}$ contains attractive 
contributions, which are the correlation terms 
$C_\mathbf{k}^{\uparrow\downarrow}$ in Eq.~\eqref{opposite-spin}. 
In order to estimate these correlations, we exploit the $SU(2)$-symmetry 
of the Mott state and the Lieb theorem \cite{Lieb89} which states that 
$\hat{\mathbf{S}}^2|\mathrm{Mott}\rangle=0$.
Then, neglecting correlations beyond neighboring sites implies 
$C_\mathbf{k}^{\uparrow\downarrow}\approx T_\mathbf{k}/(16T)$
in two dimensions.

Again assuming a square lattice where $T_\mathbf{k}$ behaves as 
$\cos k_x+\cos k_y$, holon states around the minimum at 
$(k_x,k_y)=(\pi,\pi)$ are filled up first.
Shifting the origin to the minimum $k_{x,y}\rightarrow k_{x,y}+\pi$,
the energies $E_\mathbf{k}$ scale quadratically for small $\mathbf{k}$. 
Then we may seek for d-wave solutions of the gap equation~\eqref{gapequ}
\begin{align}
\triangle_\mathbf{k}
=
\triangle^{\rm d}\left(\cos k_x-\cos k_y\right) 
\,,
\end{align}
which do also scale quadratically $k^2_y-k^2_x$ for small $\mathbf{k}$. 
As usual, pairing is expected to be most pronounced in the vicinity  
around the Fermi momentum $k_{\rm F}$ such that we restrict the 
integral~\eqref{gapequ} to the interval $|k-k_{\rm F}|<k_{\rm cut}$
with some cut-off $k_{\rm cut}\leq\ord(k_{\rm F})$.
Linearizing the energies $E_\mathbf{k}$ in this interval, we find 
\begin{align}
\label{gap}
\triangle^{\rm d}
=
\ord(T)\exp\left\{-\frac{32\pi}{3k_{\rm F}^4}\right\}\,.
\end{align}
Since we have re-scaled all length scales with respect to the lattice 
spacing $\ell$, the above Fermi momentum $k_{\rm F}$ is dimensionless.
Restoring physical units would correspond to the replacement 
$k_{\rm F}\to\ell k_{\rm F}$.

As in the usual BCS theory, we obtain an exponential suppression of the gap, 
but now the exponent is not inversely proportional to the coupling 
strength (because the kinetic and the interaction energy both scale 
linearly in $T$) but to the fourth power of the Fermi momentum, i.e.,  
the holon number density squared.
Two powers of $k_{\rm F}$ stem from the volume element, the other two from 
the quadratic scaling of the d-wave order parameter. 
The strong exponential suppression for small $k_{\rm F}$ might indicate 
that a certain holon number density is required to observe d-wave pairing 
\cite{Mo06,Ti00,Lee06,Ti02,Aba01}, 
but further investigations are needed to settle this issue.

Nonetheless, comparing our findings with the well-known phase 
diagram of cuprates, for example, 
we find qualitative consistency as superconductivity is 
usually associated with a region of finite holon doping at low temperatures. 
Of course, the range of applicability of the simple single-band Fermi-Hubbard 
model~\eqref{Fermi-Hubbard} must be taken into account in this regard.
This becomes even more important for the opposite case of electron doping. 
The particle-hole duality discussed in the Introduction implies that BCS states 
for doublons should exist in the same way as for holons. 
However, the asymmetry of the phase diagram of cuprates with respect to 
electron doping versus hole doping already shows that these systems do not 
display this particle-hole duality (as is also well known) and thus requires 
a description beyond the single-band Fermi-Hubbard 
model~\eqref{Fermi-Hubbard}.

\section{Conclusions} 

Via a combination of approaches, we studied the interaction between 
doublons or holons as quasi-particle excitations (i.e., charge modes)
of the Mott insulator state in the strongly interacting Fermi-Hubbard model. 
Using the hierarchy of correlations and the simple mean-field 
ansatz~\eqref{mean-field}, we derived a Boltzmann equation~\eqref{Boltzmann} 
with a scattering cross section which is quadratic in the hopping strength 
$T$ for doublons or holons of opposite spin (and zero otherwise). 

This motivates an effective interaction 
$V_{\mathbf{k}\mathbf{p}\mathbf{q}}^{\uparrow\downarrow}$
whose strength is linear in $T$ and which can be represented by an effective
Hamiltonian of the form~\eqref{effective}. 
Note that this effective Hamiltonian should be treated with special care:
First, the doublon and holon quasi-particle operators 
$\hat d_{{\bf k}s}$ and $\hat h_{{\bf k}s}$
do not satisfy the standard commutation relations.
Second, the Boltzmann equation does only contain the absolute value squared 
of the interaction strength 
$|V_{\mathbf{k}\mathbf{p}\mathbf{q}}^{\uparrow\downarrow}|^2$ and thus 
does not determine its sign (attractive or repulsive) uniquely. 

Although one might use continuity arguments to demonstrate that the effective 
interaction $V_{\mathbf{k}\mathbf{p}\mathbf{q}}^{\uparrow\downarrow}$
can be attractive as well as repulsive (depending on the momenta), 
we employed strong-coupling perturbation theory to infer 
$V_{\mathbf{k}\mathbf{p}\mathbf{q}}^{\uparrow\downarrow}$ including its sign.
Inserting the simple mean-field ansatz~\eqref{mean-field}, which neglects the 
correlations between lattice sites, we indeed recover the interaction 
$V_{\mathbf{k}\mathbf{p}\mathbf{q}}^{\uparrow\downarrow}$ in the effective
Hamiltonian~\eqref{effective} and thus the Boltzmann equation~\eqref{Boltzmann}
to lowest order in $T/U$. 

These calculations motivate the following intuitive picture:
In the Mott insulator state, hopping is suppressed due to the Mott gap,
such that the tunneling probabilities scale with $T^2/U^2$. 
Inserting a holon, however, the system can lower its energy by tunneling 
-- which gives rise to the single-holon quasi-particle energies of order $T$. 
Two holons far away from each other lower the energy according to the sum of 
their quasi-particle energies.
However, if they come too close, the presence of one holon can influence 
(suppress) the tunneling of the other holon and vice versa, such that the 
energy reduction changes -- giving rise to an effective interaction. 
Obviously, starting from the Mott state (containing one particle per site), 
two holons cannot occupy the same lattice site.  

Consistent with this picture, two holons with momenta $\bf k$ and $\bf p$ 
which both lower the energy separately $T_{\bf k}<0$ and $T_{\bf p}<0$ 
would repel each other while two holons which increase the energy separately
$T_{\bf k}>0$ and $T_{\bf p}>0$ would attract each other.
The above line of argument specifically applies to holons, but the 
particle-hole duality mentioned in the Introduction implies the analogous 
behavior for doublons after the substitution $T\to-T$.
(The interaction between a doublon and a holon is discussed in Appendix \ref{Doublons and holons}.)
As a result, if two holons with momenta $\bf k$ and $\bf p$ attract each other, 
two doublons with the same momenta would repel each other and vice versa. 

For the simple example of the Fermi-Hubbard model on a square 
(Hubbard tetramer) admitting an analytic solution, we could confirm the 
above picture -- at least qualitatively.
While the energy of the Mott state~\eqref{Mott-zero} scales quadratically
$\ord(T^2/U)$, the eigen-energies of the states corresponding to one and 
two holons are linear in $T$ to lowest order. 
Furthermore, the lowest (highest) two-holon eigen-energies cannot be written 
as a sum of two one-holon eigen-energies, indicating an effective 
repulsion (attraction). 

Going beyond the simple mean-field ansatz~\eqref{mean-field} and taking 
spin correlations between the lattice sites into account, we obtain 
corrections to the quasi-particle energies $E_{\bf k}$ as well as to 
their effective interaction.
For example, these correlations do also lead to an 
interaction~\eqref{same-spin} between holons of the same spin. 
Furthermore, for two holons of opposite spin, the effective interaction
$V_{\mathbf{k}\mathbf{p}\mathbf{q}}^{\uparrow\downarrow}$,
which is repulsive for low-energy holons, does also acquire attractive
corrections~\eqref{opposite-spin} due to the spin correlations. 
As an intuitive picture, the fact that two holons cannot occupy the same 
lattice site leads to an effective on-site repulsion whereas the spin 
correlations can induce a finite range attraction:
If a holon with spin $\uparrow$ occupies the lattice site $\mu$, 
there must have been an electron with that spin $\uparrow$ in the Mott state 
at that lattice site $\mu$. 
Then, in the presence of (even short-ranged) anti-ferromagnetic order,
the probability for having an electron with the other spin $\downarrow$ 
in the Mott state at a neighboring lattice site $\nu$ is larger than average. 
Thus, this neighboring lattice site $\nu$ can support 
a holon with the other spin $\downarrow$. 
In addition to the on-site repulsion explained above, one can visualize 
this as a nearest-neighbor attraction.

Note that the observed scale separation between the fast frequency scale $T$ 
of the propagation and interaction of the doublons and holons on the one hand 
and the slow frequency scale $T^2/U$ of the spin fluctuations on the other 
hand allows us to approximately treat the (fast) evolution of the doublons 
and holons as taking place on a background with a fixed spin structure. 

Finally, we discussed the implications of our results for high-temperature
superconductivity. 
Using a BCS-like variational ansatz, we found that the usual s-wave pairing 
would not lower the energy (due to the effective on-site repulsion) but 
d-wave pairing could actually reduce the energy as a result of the 
nearest-neighbor attraction.
Within the effective Hamiltonian approach, we deduced a gap equation.
Its solution for the d-wave gap displays the usual non-perturbative 
structure, but in terms of the Fermi momentum of the holons instead 
of a coupling strength.

Of course, the effective interaction between doublons and/or holons has 
already been discussed in many publications, see, e.g., \cite{Aro22} 
and references therein. 
By now it is commonly expected that the spin degrees of freedom play an 
important role in that respect. 
The major points specific to the present work are: 
First, the derivation of the Boltzmann equation (based on the $1/Z$ expansion) 
displaying scattering cross sections which scale quadratically in $T$ and 
thus point to an effective interaction linear in $T$.
Second, the derivation of this effective interaction (based on the $1/U$ expansion) 
which is indeed linear in $T$ and contains attractive as well as repulsive
contributions. 
Third, the resulting gap equation whose solution is also exponentially suppressed, 
but the exponent merely contains the holon density.

\section{Outlook} 

There are many ways to generalize our results. 
As one example, we focused on the leading order (in $1/Z$ or $T/U$).
Including higher orders would lead to modifications in several places. 
For instance, the lowest-order mean-field ansatz~\eqref{Ising} 
could be modified by including small probabilities for an empty or 
doubly occupied lattice site or that this lattice site is occupied
by the ``wrong'' spin. 
In this way, the back-reaction of the quantum or thermal fluctuations 
onto the mean field can be taken into account. 
This, in turn, would change the lowest-order quasi-particle 
energies 
a bit, which corresponds to a renormalization of the involved quantities,
quite analogous to the case of weakly interacting systems, see, e.g., \cite{G63,G65,B70,B12}. 
%
%
In a similar manner, one could include higher orders in $T/U$ in 
Sec.~\ref{Perturbation Theory}.

As a somewhat related point, we considered the zero-temperature limit here.
Finite temperatures can also be taken into account in the approach based on 
the hierarchy of correlations (as it deals with density matrices),
for example via the double-time correlator, see, e.g., \cite{Q23}. 
The expected impact of finite temperatures can be discussed in terms of 
general arguments.
If the temperature is well below the typical spin energy of order $T^2/U$,
one would expect that our results are basically unaffected.
Once the temperature is above this energy scale $T^2/U$, it is expected to 
wash out the anti-ferromagnetic correlations and thus the finite-range 
attraction (responsible for d-wave pairing) is suppressed while the 
on-site repulsion remains. 
The next characteristic scale is reached when the temperature approaches 
the hopping rate $T$ leading to thermal broadening of the holon distribution
functions. 
%
%
Finally, once the temperature reaches or even exceeds the Mott gap of 
order $U$, thermal excitations in the form of real doublon-holon pairs 
change the background~\eqref{mean-field} considerably such that the 
insulating behavior of the Mott phase disappears. 

In this context, one should also remember that we took the magnetic order
of the Mott background as given, i.e., fixed. 
As explained above, the rationale behind that is the separation of scales 
between the scale $T$ of propagation and interaction of the holons and the 
characteristic scale $T^2/U$ of the spin fluctuations. 
However, a complete picture would also require a more detailed treatment 
of the spin fluctuations and the origin of the magnetic order.
For example, even though $T$ is much larger than $T^2/U$ in the 
strong-coupling limit considered here, the superconducting gap~\eqref{gap}
scales as $T\exp\{-32\pi/(3k_{\rm F}^4)\}$ and thus it could be smaller 
than $T^2/U$ for a very low density of holons. 
In this case, the spin fluctuations might even destroy superconductivity. 

Closely related to the magnetic order is the lattice structure. 
Our approach can basically be applied to quite general lattices, 
as long as they obey the usual (discrete) translational symmetries.
The pseudo-spin $\hat{\f{\eta}}$ and the anti-ferromagnetic order such as in 
Eq.~\eqref{Ising} require bi-particle lattices.
For lattices which are not bi-particle (e.g., a triangular lattice),
the anti-ferromagnetic order would be suppressed due to frustration,
but short-range anti-ferromagnetic correlations should still persist 
(although on a weaker level) and thus the finite-range attraction 
(responsible for d-wave pairing) may survive. 
Apart from the discussion of the Hubbard tetramer in 
Sec.~\ref{Hubbard tetramer} which is obviously devoted to this specific 
example, we assumed a square lattice in Sec.~\eqref{Bardeen-Cooper-Schrieffer}. 
For other lattice structures (e.g., hexagonal), one should adapt the 
Fourier components $T_{\mathbf{k}}$ accordingly, which might then alter 
the rotational symmetries of the superconducting gap. 

It should also be illuminating to compare the results of our approach with 
other methods based on a large-$Z$ expansion such as dynamical mean-field 
theory (DMFT) \cite{L00,G96,H07,K08} or its time-dependent version (t-DMFT) 
\cite{F06,A14,W14,M18}. 
As a first difference, this method usually considers a different scaling 
limit, i.e., a factor of $1/\sqrt{Z}$ instead of $1/Z$ in front of the 
hopping term in the Hamiltonian~\eqref{Fermi-Hubbard}. 
As a consequence, already the limit $Z\to\infty$ becomes non-trivial,
while we are mostly interested in the corrections of order $1/Z$ or higher. 
Furthermore, such methods which are based on the mapping to an effectively
single lattice site or a finite cluster of sites are quite suitable for 
deriving frequency-dependent quantities such as the self-energy -- 
but are less adapted to the problem considered here, where the spatial 
structures and the momentum dependence play an important role.

Note that our considerations are solely based on the Fermi-Hubbard model without 
invoking any effective descriptions (such as the $t$-$J$ model).
However, it would be interesting to generalize our findings to other model 
Hamiltonians (see, e.g., \cite{Hat92,Hua22}) and to compare the results.

\bmhead{Acknowledgments}

The authors acknowledge fruitful discussions with 
J.~Schwardt, C.~Timm and M.~Vojta.    
Funded by the Deutsche Forschungsgemeinschaft 
(DFG, German Research Foundation) -- Project-ID 278162697-- SFB 1242. 

\begin{appendices}

\section{Derivation of Boltzmann equation}
\label{Derivation of Boltzmann equation} 
In the following, the spin configuration, the lattice sites and the part of the operator split (\ref{full+empty}) are indicated by small Latin indices.
We have then for example $a=\{\mu,\uparrow,1\}$ and $b=\{\nu,\uparrow,0\}$
for the two-site correlator $\langle \hat{c}^\dagger_{\mu,\uparrow}\hat{n}_{\mu,\downarrow} 
\hat{c}_{\nu,\uparrow}(1-\hat{n}_{\nu,\downarrow})\rangle^\mathrm{corr}\equiv \langle C^{\dagger}_{\mu,\uparrow,1}C_{\nu,\uparrow,0}\rangle=G^{(2)}_{ab}$.
The two-site correlators satisfy the equation of motion
which has the schematic form
\begin{align}\label{twosite}
i\partial_t G^{(2)}_{ab}
=\sum\limits_{cd}M_{abcd}^{(22)}G^{(2)}_{cd}
+S^{(2)}_{ab}
+\sum\limits_{\alpha,cde}\left(M_{abcde}^{\alpha\,(23)}G^{\alpha\,(3)}_{cde}
-M_{bacde}^{\alpha\,(23)}\left(G^{\alpha\,(3)}_{cde}\right)^*
\right)
\end{align}
The first term in (\ref{twosite}) determines the free linear time evolution of the correlators, 
$S^{(2)}_{ab}$ contains source terms of order $1/Z$ as well as correlators which turn out to be irrelevant for the Boltzmann evolution.
Finally, the last term in equation (\ref{twosite}) links the two-site correlators to various 
three-site correlators of order $1/Z^2$.
Here the index $\alpha$ labels the various three-point correlators.
The equation (\ref{twosite}) can be Fourier transformed and rotated into the particle-hole basis, see equation (\ref{Bogoliubov}).
The interference terms between particles and holes rapidly approach a thermalised value as they oscillate
at a frequency $\sim U$.
Therefore, they do not appear as dynamical variables in the transformed equations.
The remaining dynamical variables are the quasi-particle populations $G^{(2)}_A$, which evolve according to 
\begin{align}\label{twositetrans}
i\partial_t G^{(2)}_A=\sum_{\alpha,B}M^{\alpha\,(23)}_{AB}\left(G^{\alpha\,(3)}_{BA}-\left(G^{\alpha\,(3)}_{BA}\right)^*\right)\,.
\end{align}
Here, the multi-index $A$ contains the spin-configuration, the lattice momentum $\mathbf{k}$
and the quasi-particle index.

The dynamics of the three-point correlators is governed by equations of the 
form
\begin{align}
i\partial_t G^{\alpha\,(3)}_{abc}=&\sum_{def}M^{\alpha\,(33)}_{abcdef}G_{def}^{\alpha\,(3)}
+\sum_{de}M^{\alpha\,(32)}_{abcde}G_{de}^{(2)}\nonumber\\
&+\sum_{defg}M^{\alpha\,(322)}_{abcdefg}G_{de}^{(2)}G_{fg}^{(2)}
+\sum_{defg}M^{\alpha\,(34)}_{abcdefg}G_{defg}^{(4)}\label{threesite}
\end{align}
The first term on the right hand side of equation (\ref{threesite}) describes the linear 
evolution, the second and the third term are of order $1/Z^2$ and couple to the two-site correlations.
The last term, being of order $1/Z^3$, is of central significance for the Boltzmann collision terms as it couples the three-site terms to a correlation which contains two annihilation-operators and two creation-operators.
After Fourier transform and rotation to the particle-hole basis we find from (\ref{threesite})
\begin{align}
i\partial_t G^{\alpha\,(3)}_{AB}=&M^{\alpha\,(33)}_{AB} G^{\alpha\,(3)}_{AB}+\sum_C M_{ABC}^{\alpha\,(32)}G_C^{(2)}\nonumber\\
&+M^{\alpha\, (322)}_{AB}G_A^{(2)}G_B^{(2)}
%
+\sum_{CDEF}M_{ABCDEF}^{\alpha\,(34)}G^{(4)}_{CDEF}\label{threesitetrans}
\end{align}
Finally, the equation of motion for the four-point correlator, which is of the order of $1/Z^3$, can be written schematically as follows
\begin{align}
i\partial_t G^{(4)}_{abcd}=&\sum_{efgh}M^{(44)}_{abcdefgh} G^{(4)}_{efgh}
+\sum_{\alpha,efg} M^{\alpha\,(43)}_{abcdefg}G_{efg}^{\alpha\,(3)}\nonumber\\
&+\sum_{\alpha,efghi} M^{\alpha\,(432)}_{abcdefghi}G_{efg}^{\alpha\,(3)}
G_{hi}^{(2)}
%
+\sum_{efgh} M^{(422)}_{abcdefgh}G_{ef}^{(2)}
G_{hi}^{(2)}\label{foursite}
\end{align}
Going to Fourier space, we then obtain from (\ref{foursite})
\begin{align}
i\partial_t G_{ABCD}^{(4)}=&M^{(44)}_{ABCD}G_{ABCD}^{(4)}+\sum_{\alpha,EF}M_{ABCDEF}^{\alpha\, (43)}
G^{\alpha\,(3)}_{EF}\nonumber\\
&+\sum_{\alpha,EFG}M_{ABCDEFG}^{\alpha\, (432)}
G^{\alpha\,(3)}_{EF}G_G^{(2)}
%
+\sum_{EF}M^{(422)}_{ABCDEF}G^{(2)}_EG^{(2)}_F\label{foursitetrans}
\end{align}
The differential equations (\ref{threesitetrans}) and (\ref{foursitetrans}) both contain linear terms determined by the free quasi-particle evolution and source terms which include the coupling to other correlation functions.
These differential equations are of the form 
\begin{align}
i\partial_t f(t)=\omega f(t)+q(t) 
\end{align}
and have the formal solution
\begin{align}\label{sol}
f(t)=-i\int_0^t d\tau e^{-i\omega\tau}q(t-\tau)\,. 
\end{align}
If the source term $q(t)$ varies sufficiently slowly with time, the Markov equation $q(t-\tau)\rightarrow 
q(t)$ can be applied.
As we are interested in the long-time evolution, we can extend the time-integral (\ref{sol}) to infinity and obtain the time-local expression
\begin{align}\label{solMarkov}
f(t)\approx -\frac{1}{\omega-i\epsilon}q(t)\,.
\end{align}
Along these lines the Markov solutions of (\ref{threesitetrans}) and (\ref{foursitetrans}) can be obtained.
Finally we employ (\ref{twositetrans}) which leads in the long-time limit to
\begin{align}
&\partial_t G^{(2)}_A=
-2\pi \sum_{BCD}\delta(M^{(44)}_{ABCD})S_{ABCD}(G^{(2)}_A,G^{(2)}_B,G^{(2)}_C,G^{(2)}_D)\,.\nonumber 
\end{align}
The delta-function ensures the energy-conservation and,
due to momentum conservation, all momenta which are contained in the multi-indices 
$A,B,C,D$ add up to zero.
The scattering kernel $S_{ABCD}$ is a function of the doublon- and holon distributions $G^{(2)}_A$.

\section{Doublons and holons} 
\label{Doublons and holons} 

The explicit form of the scattering kernel $S_{ABCD}$ is a rather complicated expression.
However, in the limit of strong interactions the Boltzmann dynamics simplifies 
to
\bea
\label{Boltzmann-holons}
& &\partial_t \mathfrak{h}_\mathbf{k}^\uparrow
= 
-2\pi \int\limits_{\mathbf{p}\mathbf{q}}
\delta\left(\frac{T_\mathbf{k}}{2}+\frac{T_\mathbf{p}}{2}-\frac{T_\mathbf{k+q}}{2}-\frac{T_\mathbf{p-q}}{2}\right)
\\
&&
\times 
\Big\{ 
\left(T_\mathbf{k}+T_\mathbf{p}\right)^2 
\left[
\mathfrak{h}_\mathbf{k}^\uparrow
\mathfrak{h}_\mathbf{p}^\downarrow
\left(1-\mathfrak{h}_\mathbf{k+q}^\uparrow\right)
\left(1-\mathfrak{h}_\mathbf{p-q}^\downarrow\right)
-
\mathfrak{h}_\mathbf{k+q}^\uparrow
\mathfrak{h}_\mathbf{p-q}^\downarrow
\left(1-\mathfrak{h}_\mathbf{k}^\uparrow\right)
\left(1-\mathfrak{h}_\mathbf{p}^\downarrow\right)
\right] 
\nn
&& 
\quad
+(T_\mathbf{p}-T_\mathbf{p-q}
)^2\left[
\mathfrak{h}_\mathbf{k}^\uparrow
\mathfrak{p}_\mathbf{p}^\downarrow
\left(1-\mathfrak{p}_\mathbf{k+q}^\uparrow\right)
\left(1-\mathfrak{h}_\mathbf{p-q}^\downarrow\right)
-
\mathfrak{p}_\mathbf{k+q}^\uparrow
\mathfrak{h}_\mathbf{p-q}^\downarrow
\left(1-\mathfrak{h}_\mathbf{k}^\uparrow\right)
\left(1-\mathfrak{p}_\mathbf{p}^\downarrow\right)
\right]\nn
&&
\quad+(T_\mathbf{p}-T_\mathbf{k+q})^2
\left[
\mathfrak{h}_\mathbf{k}^\uparrow
\mathfrak{p}_\mathbf{p}^\downarrow
\left(1-\mathfrak{h}_\mathbf{k+q}^\uparrow\right)
\left(1-\mathfrak{p}_\mathbf{p-q}^\downarrow\right)
-
\mathfrak{h}_\mathbf{k+q}^\uparrow
\mathfrak{p}_\mathbf{p-q}^\downarrow
\left(1-\mathfrak{h}_\mathbf{k}^\uparrow\right)
\left(1-\mathfrak{p}_\mathbf{p}^\downarrow\right)
\right]
\Big\}\nonumber
\ea
and
\bea
\label{Boltzmann-doublons}
& &\partial_t \mathfrak{p}_\mathbf{k}^\uparrow
= 
-2\pi \int\limits_{\mathbf{p}\mathbf{q}}
\delta\left(\frac{T_\mathbf{k}}{2}+\frac{T_\mathbf{p}}{2}-\frac{T_\mathbf{k+q}}{2}-\frac{T_\mathbf{p-q}}{2}\right)
\\
&&
\times 
\Big\{ 
\left(T_\mathbf{k}+T_\mathbf{p}\right)^2 
\left[
\mathfrak{p}_\mathbf{k}^\uparrow
\mathfrak{p}_\mathbf{p}^\downarrow
\left(1-\mathfrak{p}_\mathbf{k+q}^\uparrow\right)
\left(1-\mathfrak{p}_\mathbf{p-q}^\downarrow\right)
-
\mathfrak{p}_\mathbf{k+q}^\uparrow
\mathfrak{p}_\mathbf{p-q}^\downarrow
\left(1-\mathfrak{p}_\mathbf{k}^\uparrow\right)
\left(1-\mathfrak{p}_\mathbf{p}^\downarrow\right)
\right] 
\nn
&& 
\quad
+(T_\mathbf{p}-T_\mathbf{p-q}
)^2\left[
\mathfrak{p}_\mathbf{k}^\uparrow
\mathfrak{h}_\mathbf{p}^\downarrow
\left(1-\mathfrak{h}_\mathbf{k+q}^\uparrow\right)
\left(1-\mathfrak{p}_\mathbf{p-q}^\downarrow\right)
-
\mathfrak{h}_\mathbf{k+q}^\uparrow
\mathfrak{p}_\mathbf{p-q}^\downarrow
\left(1-\mathfrak{p}_\mathbf{k}^\uparrow\right)
\left(1-\mathfrak{h}_\mathbf{p}^\downarrow\right)
\right]\nn
&&
\quad+(T_\mathbf{p}-T_\mathbf{k+q})^2
\left[
\mathfrak{p}_\mathbf{k}^\uparrow
\mathfrak{h}_\mathbf{p}^\downarrow
\left(1-\mathfrak{p}_\mathbf{k+q}^\uparrow\right)
\left(1-\mathfrak{h}_\mathbf{p-q}^\downarrow\right)
-
\mathfrak{p}_\mathbf{k+q}^\uparrow
\mathfrak{h}_\mathbf{p-q}^\downarrow
\left(1-\mathfrak{p}_\mathbf{k}^\uparrow\right)
\left(1-\mathfrak{h}_\mathbf{p}^\downarrow\right)
\right]
\Big\}
\,.\nonumber
\ea
The simple Boltzmann form allows to construct an effective 
Hamiltonian from which (\ref{Boltzmann-holons}) and (\ref{Boltzmann-doublons})
can be recovered using leading order perturbation theory together 
with the usual Markov approximation,
\begin{align}\label{efftotal}
&\hat{H}_\mathrm{eff}=\sum_{s}\int_\mathbf{k}\left(E^-_\mathbf{k}
\hat{a}^{\dagger}_{\mathbf{k},s}\hat{a}^{}_{\mathbf{k},s}+E^+_\mathbf{k}
\hat{b}^{\dagger}_{\mathbf{k},s}\hat{b}^{}_{\mathbf{k},s}\right)\\
&+\frac{1}{2}\int_{\mathbf{kpq}}(T_{\mathbf{p-q}}+T_{\mathbf{p}}+T_{\mathbf{k+q}}+T_{\mathbf{k}})\left[\hat{a}^{\dagger }_{\mathbf{k+q},\uparrow}\hat{a}_{\mathbf{k},\uparrow}
\hat{a}^{\dagger }_{\mathbf{p-q},\downarrow}\hat{a}_{\mathbf{p},\downarrow}-
\hat{b}^{\dagger}_{\mathbf{k+q},\uparrow}\hat{b}_{\mathbf{k},\uparrow}
\hat{b}^{\dagger}_{\mathbf{p-q},\downarrow}\hat{b}_{\mathbf{p},\downarrow}
\right]\nonumber\\
&
+\frac{1}{2}\int_{\mathbf{kpq}}(-T_{\mathbf{p-q}}-T_{\mathbf{p}}+T_{\mathbf{k+q}}+T_{\mathbf{k}})\left[\hat{a}^{\dagger }_{\mathbf{k+q},\uparrow}\hat{a}_{\mathbf{k},\uparrow}
\hat{b}^{\dagger}_{\mathbf{p-q},\downarrow}\hat{b}_{\mathbf{p},\downarrow}-
\hat{b}^{\dagger}_{\mathbf{k+q},\uparrow}\hat{b}_{\mathbf{k},\uparrow}
\hat{a}^{\dagger}_{\mathbf{p-q},\downarrow}\hat{a}_{\mathbf{p},\downarrow}
\right]\nonumber\\
&
+\frac{1}{2}\int_{\mathbf{kpq}}(-T_{\mathbf{p-q}}+T_{\mathbf{p}}+T_{\mathbf{k+q}}-T_{\mathbf{k}})\left[\hat{a}^{\dagger}_{\mathbf{k+q},\uparrow}\hat{b}_{\mathbf{k},\uparrow}
\hat{b}^{\dagger}_{\mathbf{p-q},\downarrow}\hat{a}_{\mathbf{p},\downarrow}-
\hat{b}^{\dagger}_{\mathbf{k+q},\uparrow}\hat{a}_{\mathbf{k},\uparrow}
\hat{a}^{\dagger}_{\mathbf{p-q},\downarrow}\hat{b}_{\mathbf{p},\downarrow}
\right]\,.\nonumber
\end{align}
Here the operators $\hat{a}_{\mathbf{k},s}$ and $\hat{b}_{\mathbf{k},s}$
are the annihilation operators for holons and doublons, respectively.
As shown in section (\ref{twoholonstate}), the effective interaction potential $V^{\uparrow\downarrow,\mathfrak{h}\mathfrak{h}}_{\mathbf{kpq}} $ for holon-holon scattering can also be justified from leading order perturbation theory involving 
two-holon states.
Similarly, taking as initial and final state two-doublon wave packets, one finds the effective interaction potential $V^{\uparrow\downarrow,\mathfrak{p}\mathfrak{p}}_{\mathbf{kpq}}=-V^{\uparrow\downarrow,\mathfrak{h}\mathfrak{h}}_{\mathbf{kpq}}$
where the minus sign originates from the particle-hole symmetry.
The holon-doublon scattering can be justified from holon-doublon wavepackets of the form
\begin{align}
|\Psi_{\mathrm{in}(\mathrm{out})}\rangle_0=\hat{e}^\dagger_{\mathbf{k}\uparrow} \hat{f}^\dagger_{\mathbf{p}\downarrow}|\mathrm{Mott}\rangle_0=\mathcal{N}_{\mathbf{k}\mathbf{p}}^{\uparrow\downarrow} \sum_{\alpha\beta}\hat{c}_{\alpha\uparrow}\hat{c}^\dagger_{\beta\downarrow}(1-\delta_{\alpha\beta})|\mathrm{Mott}\rangle_0\, e^{i\mathbf{k}\cdot \mathbf{r}_\alpha+i\mathbf{p} \cdot\mathbf{r}_\beta}\,,
\end{align}
which is consistent with the fourth line of the effective Hamiltonian (\ref{efftotal}).
The exchange interaction in the last line of equation (\ref{efftotal}) was choosen in order to achieve full consistency with the Boltzmann equations (\ref{Boltzmann-holons}) and (\ref{Boltzmann-doublons}).


\section{Pseudo-spin}\label{APP:pseudospin} 

For bi-partite lattices it is possible to find an ordering of the lattice sites $\mu$ such that the parity $(-1)^\mu$ is always 
opposite for lattice neighbors, i.e., $(-1)^{\mu+\nu} = -1$ when $T_{\mu\nu}\neq 0$.
One can -- analogous to the raising and lowering operators $\hat S^\pm = \hat S^x \pm \ii \hat S^y$ that one defines for the spin~\eqref{spin} -- introduce the pseudo-spin ladder operators
\begin{align}
\hat\eta &=\sum_\mu (-1)^\mu \hat c_{\mu\uparrow}\hat c_{\mu\downarrow}
\,,\quad 
\hat\eta_z=\frac12\left(\hat N-N_{\rm lattice}\right)\,,  
\end{align}
where $N_{\rm lattice}$ denotes the number of lattice sites. 
Analogous to the spin ladder operators, these obey the relations $[\hat\eta,\hat\eta^\dagger]=-2\hat\eta_z$, $[\hat\eta_z, \hat\eta]=-\hat\eta$, and $[\hat\eta_z, \hat\eta^\dagger]=+\hat\eta^\dagger$.
The square of the total pseudo-spin can then be written as
\begin{align}\label{EQ:pseudospin}
\hat{\f{\eta}}^2
=
\frac12\left(\hat\eta\hat\eta^\dagger+\hat\eta^\dagger\hat\eta\right)
+\hat\eta_z^2
\,.
\end{align}
For a bi-partite lattice one finds that it commutes with $\hat H$, $\hat N$, $\hat{\f{S}}$, and $\hat{\f{S}}^2$.
Unlike the spin, $\hat \eta$ and $\hat \eta^\dagger$ (or $\hat \eta^x = \frac{1}{2} (\hat \eta+\hat \eta^\dagger)$ and $\hat \eta^y = \frac{1}{2\ii} (\hat\eta-\hat\eta^\dagger)$) do not separately commute with $\hat H$.
Only in the special case with finite on-site energies $T_{\mu\mu} = \epsilon = Z U/2$ one finds that $\hat \eta$, $\hat \eta^\dagger$ also commute with $\hat H$.

\end{appendices}



\end{document}